\documentclass[fleqn,10pt,twoside]{article}

\usepackage[english] {babel}

\usepackage {graphicx}
\usepackage {graphics}
\usepackage {floatflt}
\usepackage {latexsym}
\usepackage {caption3} 
\usepackage[labelsep=period]{caption}[2007/11/04 v.3.1e]
\usepackage {amssymb}

\def\noi{\noindent}

\renewcommand{\thesubsubsection}%
        {\arabic{section}.\arabic{subsection}.\arabic{subsubsection}.}

\newcommand{\heads}[2]{\markboth{\protect\small\it #1}{\protect\small\it #2}}

\newcommand{\Arthead}[5]{ \setcounter{page}{#4}\thispagestyle{empty}\noi
    \unitlength=1pt \begin{picture}(500,40)

        \put(0,58){\shortstack[l]{\small\it ISSN 0202-2893, Gravitation \& Cosmology,
                          #1, Vol.#2, No. #3, pp. #4--#5    
\footnotesize\copyright \ Pleiades Publishing Ltd., 2011.} }

    \end{picture}
	 }     		
\def\prepno#1#2
    {\thispagestyle{empty}
    \noi    \unitlength=1mm
    	\begin{picture}(178,10)
            \put(177,15){\llap{\bf #1}}
            \put(177,10){\llap{\small\rm #2}}
        \end{picture}
          }             

\newcommand{\Title}[1]{\noi {\uppercase{\Large #1}}     }
\newcommand{\Author}[2]{\noi{\large\bf #1}\\[2ex]\noindent{\it #2}   }

\newcommand{\Abstract}[1]{\vskip 2mm \begin{center}
        \parbox{16.4cm}{\small\noi #1} \end{center}\medskip}

\newcommand{\foom}[1]{\protect\footnotemark[#1]}

\newcommand{\email}[2]{\footnotetext[#1]{e-mail: #2}
		\addtocounter{footnote}{1}}

\heads{A.M.Baranov}
      {THE CATASTROPHE THEORY, PETROV'S ALGEBRAIC CLASSIFICATION ...} %

\begin{document}
\twocolumn 
[
\Arthead{2011}{17}{2}{170}{172.}

\vspace{0.5cm}
\begin{center}
\Title{The catastrophe theory, Petrov's algebraic classification 

\vspace{0.2cm}
   and gravitation phase transitions\foom2}
\end{center}

\begin{center}
\vspace{.5cm}
   \Author{A.M.Baranov\foom1}   
{\it Dep. of Theoretical Physics, Siberian Federal University,
Svobodny prosp. 79, Krasnoyarsk 660041, Russia}

{\it Received October 4, 2010}
\end{center}

\Abstract
{A close relationship is demonstrated between the catastrophe theory, Petrov's algebraic classification of the gravitational fields and the theory of phase transitions.}

{\bf DOI:} 10.1134/S0202289311020058
\vspace{1.0cm}
] 

\email 1 {alex\_m\_bar@mail.ru; AMBaranov@sfu-kras.ru}
\footnotetext[2]{Talk given at the International Conference RUDN-10, 
 June 28--July 3, 2010, PFUR, Moscow.}

\section{Introduction}

\indent

Petrov's classification ([1],[2]) is an algebraic classification of the 4D space-time on the one hand but, on the other hand, it is an algebraic classification of gravitational fields. 

\vspace{.1cm}
This classification is that of the Riemann curvature tensor or the conformal curvature (the Weyl tensor).

\vspace{.1cm}
The original version of this classification was created by Petrov in the matrix representation in 3D Euclidean complex space. The main idea of the classification is solution of an eigenvalue problem of the Weyl traceless symmetric $3 \times 3$ complex matrices. The classification is constructed on the basis of the eigenvalues and eigenvalue vectors of the Weyl matrices. 

\vspace{.1cm}
The set of Weyl matrices is decomposed onto classes (types) of $3 \times 3$ Weyl matrices depending on the eigenvalues and eigenvalue vectors. There are seven types of such Weyl matrices: {\it\bfseries I, Ia, II, D, III, N, 0}. Each type of Weyl matrix corresponds to a specific type of gravitational field. 

\vspace{.1cm}
In particular, the gravitational field of {\it\bfseries N} type is an analog of the field of a plane electromagnetic wave (the transverse-transverse gravitational wave). The type {\it\bfseries III} fields are also wave fields, but with a longitudinal component. The analog of the Coulomb field is a type {\it\bfseries D} field. The {\it\bfseries II} type is a combination of the type {\it\bfseries D} and {\it\bfseries N} fields. In a weak approximation, such a field can be considered as a superposition of the Coulomb and plane wave gravitational fields. Type {\it\bfseries Ia} fields are an analog of electromagnetic stationary waves. The canonical {\it\bfseries Ia} type of Weyl matrix can be written in the form $W_{{\bf Ia}} = diag(0,1, -1).$ This algebraic type is connected with a gravitational field of head-on scattering of weak gravitational type {\it\bfseries N} waves. A generalization of Minkowski space-time is type {\it\bfseries 0} field with a conformally flat metric. Gravitational fields which have not symmetry belong to the most general type {\it\bfseries I}.

\vspace{.3cm}
\section{Catastrophe theory and algebraic classification of 
gravitational fields}

\vspace{.1cm}
The Petrov algebraic classification of space-time is close-\linebreak 
ly connected with solutions of the cubic eigenvalues equation ([1],[2])

$$
\lambda^3 + p\lambda + q =0,
\eqno{(1)}
$$
where the set of values $\lambda$ is the set of eigenvalues, and the parameters $p$ and $q$ are the control parameters on the ($p,q$) plane, respectively (see Fig.1).

This equation can be considered to be an existence condition for critical points ofthe ''potential'' function 

$$
U={\lambda^{4}}/4+p {\lambda^{2}}/{2+q \lambda}.
\eqno{(2)}
$$

Such a ''potential'' function $U$ describes a cusp catastrophe [3]. 

\vspace{.1cm}

Eq.(1) has three roots. The separatrix equation, or the Neile equation of a half-cubic parabola,

$$
Q = (p/3)^3 + (q/2)^2 = 0,
\eqno{(3)}
$$
leads to equality of two of three roots, where $Q$ is the discriminant of the cubic equation.

\vspace{.1cm}
Thus in this case we have types {\it\bfseries D} and {\it\bfseries II} according to the algebraic classification of space-time. Elimination of variable $\lambda$ from the two equations
$$
d^2 U/d{\lambda^2}= 3 \lambda^2 +p =0;
$$
$$
d^3 U/d{\lambda^3}= 6 \lambda^2 =0.
\eqno{(4)}
$$
leads to the requirement $q=0$ (degeneracy of the Weyl matrix). This means that 
$\lambda_1 = 0;\; \lambda_2 = -\lambda_3 =\sqrt{p}.$ Therefore there is a one dimension Maxwell set on the parameter plane $(p,q).$ Solution of Eq.(4) gives the cusp point $p=q=0$ (the thrice degenerate point). 


\begin{figure}[t,h]
\center{\includegraphics[width=0.95 \linewidth]{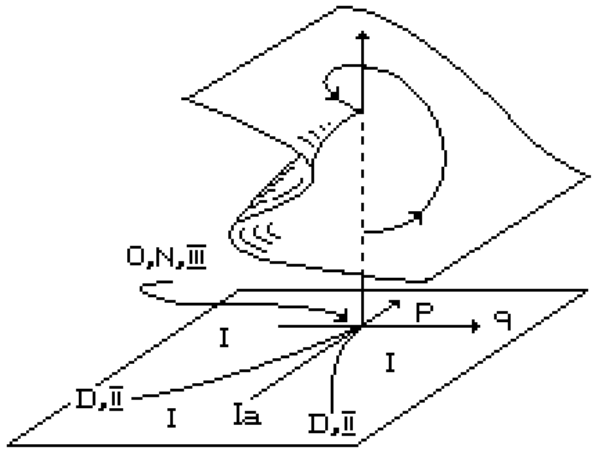}}
\caption{The cusp catastrophe's surface and its projection onto the plane of the control parameters $p$ and $q$.}
\end{figure}



\vspace{.1cm}
On the other hand, Petrov's algebraic classification is the algebraic classification of gravitational fields. On the plane ($p,q$) the Petrov algebraic types are marked as sets {\it\bfseries I,Ia, II, D} and the degenerate points as {\it\bfseries III, N, 0}. These algebraic types correspond to "the phase states" of gravitational fields. For example, types {\it\bfseries D} and {\it\bfseries II} correspond to the set of points of half-cubic parabola (the Neile curve). Type {\it\bfseries Ia} corresponds to the Maxwell manifold where there are the first-order phase transitions and the function $U$ is structurally unstable. At the cusp point we have a point of a second-order phase transition. There are three degenerate algebraic types: {\it\bfseries III, N, 0}. A transition into type {\it\bfseries 0} (the conformally flat algebraic type) is a transition into the "most symmetric phase".

\vspace{.1cm}
If we consider the parameter $p$ as an analog of a temperature, then the first-order and second-order derivatives of the function $U$ behave as the entropy and the thermal capacity, respectively. Then we have a physical interpretation of jumps of ranks of the Weyl matrices as jumps of the corresponding thermal capacity in the cusp. It is a second-order phase transition in the space of algebraic types, i.e. a phase transition between the phase states (algebraic types) of gravitational fields.

\vspace{.1cm}
It is known that all algebraic types of gravitational fields (except type {\it\bfseries I}) are very unstable under small perturbations by other algebraic types (other gravitational fields) as can be seen in [4]. From the point of view the catastrophe theory, it is clear that a change in the type of gravitational field (of the Petrov algebraic type) is connected with phase transitions of specific order [5]. 

\vspace{0.4cm}
\section{Some examples of gravitational phase transitions }

\vspace{.1cm}
The first example is the limiting field of a rapidly moving particle as its velocity tends to the speed of light and the rest mass tends to zero. Such a limiting procedure is called the lightlike limit ([6],[7]). 

\vspace{.1cm}
Using this limiting procedure for Schwarzschild, Kerr and NUT particles leads to lightlike particles: lightons and helixons ([6],[7]). Under such a procedure, the gravitational fields of particle-like sources with rest masses undergo a sudden second-order phase transition. The algebraic type {\it\bfseries D} of the gravitational fields of particle-like sources passes on into type {\it\bfseries N} or {\it\bfseries III} (the wave algebraic types). 

\vspace{.1cm}
Hence the lightlike limit correspond to the cusp catastrophe at the level of Weyl's matrix with a change of the gravitational field symmetry of such a source (a change of the Killing vectors) ([6],[7]).

\vspace{.1cm}
The second example is connected with a change of the algebraic type of the gravitational field at the center of a spherical fluid ball. It is shown in [8] that at the center of a ball and in its neighborhood we have the algebraic type {\it{\bfseries 0}} while in the remaining part of the ball there is type {\it\bfseries D} of the gravitational field. In other words, at the center and its neighborhood there is a second-order phase transition  as a change of the algebraic type of space-time [8].

\vspace{.1cm}
\section{Summary}

\vspace{.1cm}
The Petrov algebraic classification is the classification of admissible gravitational fields of different types. Transitions between types of this classification are analogs of phase transitions in condensed matter. The catastrophe theory is a generalization of the Landau phase transition theory. In this paper, a relationship between this theory and the Petrov classification is shown.

\end{document}